\documentstyle[aps,prb,twocolumn,graphics,epsfig,psfig]{revtex}

\def\eps{\epsilon}
\def\ep{\epsilon_{\rm m}}
\def\o{\nu_0}
\def\cl{c_{\rm light}}

\begin{document}

\title{Transmission Studies of Left-handed Materials }
\author{P. Marko{\v{s}}\thanks{Permanent address:
Institute of Physics, Slovak Academy of Sciences,
D{\'{u}}bravsk{\'{a}} cesta 9, 842 28 Bratislava, Slovakia}
~and C. M. Soukoulis\\
Ames Laboratory and Department of Physics and Astronomy,
Iowa State University, Ames, Iowa 50011}
\maketitle

\begin{abstract}
Left-handed materials are studied numerically using an improved
version of the transfer-matrix
method. The transmission, reflection, the phase of the reflection  and
the absorption are calculated and compared with experiments for both
single split-ring resonators (SRR) with negative permeability and  left-handed
materials (LHMs),
which have both the permittivity $\eps$ and permeability $\mu$ negative.
Our results suggest ways of positively identifying materials that have
both $\eps$ and $\mu$ negative, from materials that have either
$\mu$ or $\eps$ negative.
\end{abstract}

\bigskip
\bigskip

   	Photonic band gap (PBG) structures were originally introduced
to control the electromagnetic wave propagation of matter \cite{1,2}.
Not only dielectric \cite{1,2} but metallic structures \cite{3,4,5} were
proposed for applications in the microwave and infrared regions.
Very recently, a new area of research, called left-handed materials
(LHM), has been introduced by Pendry et al. \cite{6a,6b} and Smith et al.
\cite{7}.
LHM are by definition composites, whose properties are not determined
by the fundamental physical properties of their constituents but by
the shape and distribution of specific patterns included in them.
Thus, for certain patterns and distribution, the measured effective
permittivity $\eps_{\rm eff}$ and
the effective permeability $\mu_{\rm eff}$ can
be made to be less than zero.  In such materials, the phase and group
velocity of an electromagnetic wave propagate in opposite directions
giving rise to a number of novel properties \cite{8}.  This behavior has
been called ``left-handedness", a term first used by Veselago \cite{9} over
thirty years ago, to describe the fact that the electric field,
magnetic intensity and propagation vector are related by a
left-handed rule.

	By combining a 2D array of split-ring resonators (SRRs) with
a 2D array of wires, Smith et al. \cite{7} demonstrated for the first time
the existence of left-handed materials.  Pendry et al. \cite{6b} has
suggested that an array of SRRs give an effective $\mu_{\rm eff}$, which
can be
negative close to its resonance frequency.  It is also well known \cite{3,6a}
that an array of metallic wires behaves like a high-pass filter,
which means that the effective dielectric constant is negative at low
frequencies.  Very recently, Shelby et al. \cite{10} demonstrated
experimentally that the index of refraction $n$ is negative for a LHM.
Also, Pendry \cite{11} has suggested that a LHM with negative $n$ can
make a perfect lens.

	In this Letter, we present systematic numerical results for
the transmission, reflection and absorption properties of left-handed
materials.  An improved version of the transfer-matrix method \cite{12}
(TMM) is used.
Qualitative agreement with experiments \cite{7,14} is obtained, and new
predictions about the absorption and the phase of the reflection,
to be checked experimentally, are presented.

The transfer-matrix method \cite{12} is used to calculate the EM
transmission and reflection of LHMs.  In the TMM, the total volume of
the system is divided into small cells and fields in each cell are
coupled to those in the neighboring cell.  Then, the transfer matrix
is defined by relating the incident fields on one side of the LHM
structure with the outgoing fields on the other side.  The main
advantage of the TMM is the calculation of the transmission and
reflection coefficients for EM waves of various frequencies incident
on a finite thickness slab of a LHM.  In this case, the material is
assumed to be periodic in the directions parallel to the interfaces.
The TMM has been previously \cite{1} applied in studies of metallic
structures \cite{3,4,5,13}, as well as in dielectric structures.  In all
these examples, the agreement between theoretical predictions and
experimental measurements was very good \cite{1}.

Following the algorithm described in \cite{12}, we develop the new version
of the transfer-matrix code. The main change is the faster
normalization of the transmitted waves in the calculation of the transmission
coefficient through the structure.
  For the structures described in Figure 1,
the typical CPU time necessary to calculate transmission for a given
frequency is approximately  55 sec/unit cell  on an alpha workstation
\cite{alpha}.
Most of the CPU time
is spent in the  normalization of the actual numerical data.
As our structure is highly non-homogeneous (permittivity of systems  varies
in three or more orders from site to site), we have to normalize the numerical
data after 3-4 steps.
For studies of more homogeneous samples, the CPU
time could be reduced by a factor of 2 to 5.

	The TMM has been used to simulate the reflection and
transmission from an array of square split-ring resonators (SRRs).
Figure 1a shows a diagram of a single square SRR of the type used for
our simulations and for experiments \cite{14}. In our case we have
$c=d=g=0.33$ mm and the size of SRR $w=3$ mm. Figure 1b shows a three
dimensional realization
of the actual LHM that we have simulated.

	Figure 2 presents the results of the transmission versus
frequency for the split ring resonators alone, and of the LHM, which
consists of the SRRs with metallic wires placed uniformly between the
SRRs.  The square array of metal wires alone behaves as a high pass
filter with a cutoff frequency $\nu_c$ = 19 GHz.  The cutoff frequency
$\nu_c$ of the metallic wires is given by $\nu_c=\cl/2d\sqrt{\eps_0}$,
where $d$ is the distance between the wires, $\cl$ is the velocity of light
in the air and $\eps_0$ is the dielectric constant of the
background.  In our case $d = 5$ mm  and $\eps_0 = 1$.  The
cutoff frequency is independent of either value of Re $\ep$ or
Im $\ep$ of the metal, provided that either $|{\rm Re~} \ep| > 1000$
or ${\rm Im~} \ep > 1000$.  The dot-dashed curve is that of the SRR array with
$d =5$ mm.  By adding wires uniformly between the SRRs, a pass band
occurs where both $\mu_{\rm eff}$ and $\eps_{\rm eff}$ are negative (solid
line).
The transmitted power of the LHM is very high (close to one), because
no Im $\ep$  is taken for  the metal.
We also studied  more realistic cases with non-zero Im $\ep$.
We found that the resonance frequency $\nu_0$ increases as $|\ep|$
increases and saturates
to a value of  $\o\approx 8.5$ as $|\ep|\to\infty$.
This is clearly shown in Fig. 3 where we plot $\o$ versus the magnitude of
$\ep=\eps_r+i\eps_i$. The resonance frequency is also sensitive to the
permittivity of the board \ that the SRR is
lying and of the embedding medium. We have found that $\o$ drops as the
dielectric constant of the embedding media
$\eps_a$ increases. $\o$ drops from its value of 8 GHz for $\eps_a=1$  to a
value of 6 GHz when $\eps_a=2.0$.
  The value $\o$ increases by decreasing  the value of the permittivity
$\eps_b$ of the dielectric board. $\o$
increases to a value of 9 GHz from 8 GHz when $\eps_b$ drops to a value of
1.4 from its original value of
$\eps_b=3.4$.

The results presented in Fig. 2 are done by assuming that the permittivity of
the metal $\ep$ had a large negative value. As one can see from Fig. 2,
the value of the transmission for the LHM is very close to one. This value
drops considerable if one uses
the fact that $\ep$ for the metal has a complex permittivity. In Fig. 4a we
present results similar to that in Fig.
2, with the only difference being the value of the Im $\ep$. Notice that
the position of the resonance peak is at the
same value at around 7.8 GHz. Due to the non-zero Im $\ep$ there is
considerable drop of the peak of the transmission
for the LHM. We have not done a systematic effort to fit the experimental
results of Smith et al. \cite{7}.

In Fig. 4b, we present  new results of the frequency dependence of the
absorption for either the SRR or  the LHM. For the arrays of SRRs alone, we
find that absorption has two peaks, shown as dashed lines in Fig. 4b, while
the LHM has only one absorption peak (shown as a solid line in Fig. 4b) at
the center of its
resonance frequency.
This might be a reasonable way of identifying materials that are
left-handed and have both their
$\eps_{\rm eff}$ and $\mu_{\rm eff}$ negative. On the inset of Fig. 4b
we present the  frequency dependence of the absorption as they were found
  for a simple
model, using  \cite{6a,6b}  the following forms of the frequency dependence
of the
effective permittivity and permeability:
\begin{equation}\label{eps_eff}
\eps_{\rm eff}(\nu)=1-\frac{\nu_p^2}{\nu^2+i\nu\gamma}
\end{equation}
where $\nu_p$ is the plasma frequency or cut-off frequency $\nu_c$
of the wires, $\nu_c=\nu_p\approx 19$ GHz in our case. A negative $\mu_{\rm
eff}$
accounts for the deep in the transmission in SRRs and its form \cite{6a,6b} is
given by
\begin{equation}\label{mu_eff}
\mu_{\rm eff}(\nu)=1-\frac{F\nu^2}{\nu^2-\nu_0^2+i\nu\Gamma}
\end{equation}
where $F$ is the filling factor which is in our case close to 0.3
and $\nu_0\approx 7.9$ GHz is the resonance frequency of the SRR. We used that
$\gamma=\Gamma=0.2$ GHz.
There is very good agreement between the solution of present model and
numerical simulations of real structures.

In Fig. 5 we present the results of the phase of the reflection for both
the LHM and the array of SRRs.
  Notice that there is a substantial difference in their characteristics,
which can help us identifying meta-materials
that either have negative $\mu_{\rm eff}$ or both negative $\mu_{\rm eff}$
and $\eps_{\rm eff}$.

Finally, in Fig. 6 we present results of our numerical experiments
with various vacancies in SRR and LHM \cite{cavity}. First, we remove one SRR
from the structure.  In PBG materials, where a gap is created by the
interference effect of propagating waves, such a vacancy causes the
appearance of sharp resonance frequency inside a gap. As it is seen
in Fig 6a, the absence of one split ring resonator does not influences
neither the position of the gap nor its form. Only the amplitude of 
the transmission decreases.
This proves that the
origin of the gap is not due to the interference effect but, as we assumed,
in the negative effective permeability of the structure.
Fig. 6 also shows how the transmission peak changes when one SRR
or one wire is removed from the periodic structure of a LHM.
In the absence of one SRR, the transmission peak decreases by more than one
order of magnitude and becomes narrower (see Fig. 6a), while the wire 
vacancy creates an ``inverse
resonance" inside the transmission peak (see Fig. 6b).

\medskip

We used an improved transfer-matrix method for numerical studies of
complex meta-materials. Our  numerical data of  transmission
confirms  the presence of a resonant gap, in agreement with
theoretical predictions and experiments. Our method enables us to
analyze also the reflection and absorption of the light and also the
change of the phase of the reflected field.
Obtained numerical data are consistent with a simple homogeneous model defined
by effective permittivity and permeability. This supports our belief that
for the frequencies in the resonance gap our structure possesses
negative effective $\mu_{\rm eff}$ and negative effective $\eps_{\rm eff}$.
Although  our method does not allow us to
vary the size  parameters of SRR continuously, we are able to predict,
at least qualitatively,
how the position of resonance gap depends on various
parameters of the system.  We believe that generalization of our numerical
method to non-homogeneous discretization of space
will enable us to analyze  more realistic structures.

\medskip

We thank D.R. Smith and I. El-Kady for fruitful discussions.
Ames Laboratory is operated for the U.S.Department of Energy by Iowa
State University under Contract No. W-7405-Eng-82. This work was supported by
the Director of Energy Research, Office of Basic Science. P.M. thanks
Ames Laboratory for its hospitality and support and Slovak Grant Agency
for financial support.

\newpage

\begin{figure}[b!]
\centerline{\epsfig{file=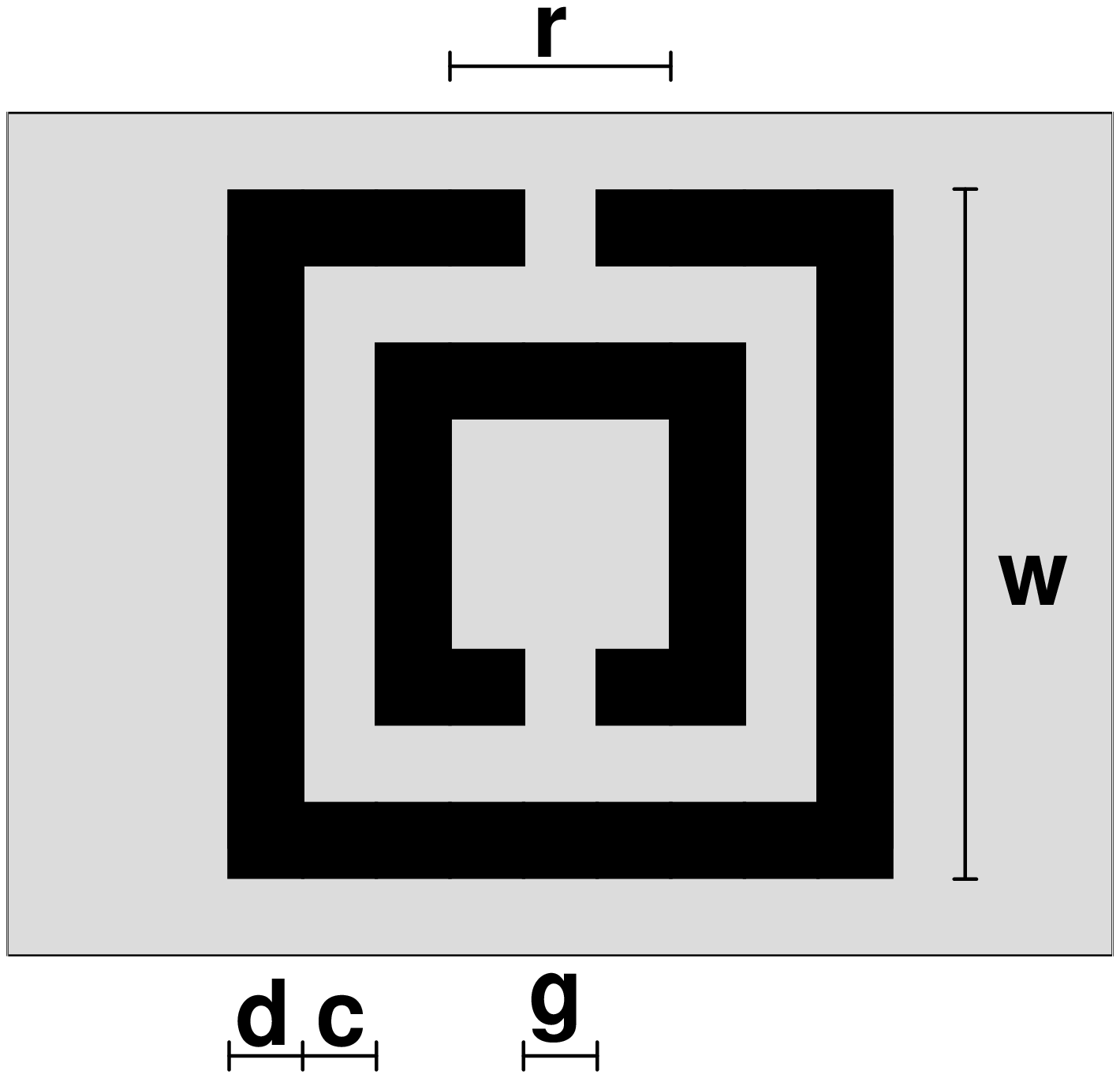,width=6cm}}
\vspace*{2mm}
\epsfig{file=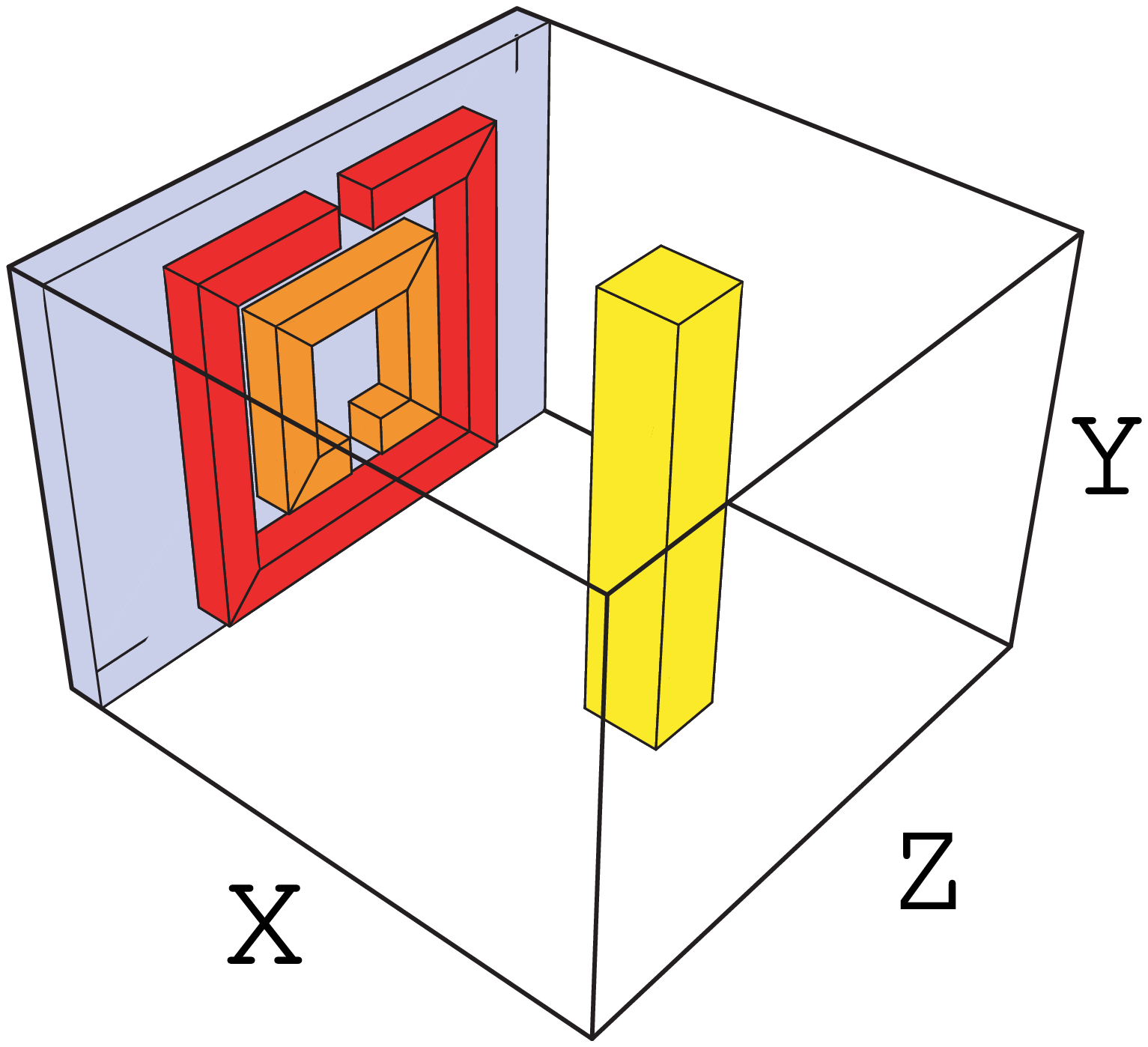,width=8cm}
\vspace*{3mm}
\caption{  (a) SRR consists from two splitted metallic rings located on a
dielectric board. We approximate the rings by squares  of size $w$.
Other  parameters, which influence the resonance frequency $\o$ are
the ring  width $c$, the radial gap $d$,
and the azimuthal gap $g$.
(b) Configuration of  the unit cell
of a LHM. Light propagates  along the $z$ axis and the magnetic intensity
is parallel to the SRR axis, which is the $x$-axis.
We assume  periodic boundary conditions  in $x$ and $y$ directions.
In the numerical simulations the unit cell is $5\times 3.63\times 5$ mm  and
is divided to $N_x\times N_y\times N_z=15\times 11\times 15$
small cells. This defines the length unit =0.33 mm.
Number of unit cells in $z$ direction
varies from 6 to 20, and is 9 for most of the cases  presented here.}
\end{figure}

\begin{figure}[t]
\epsfig{file=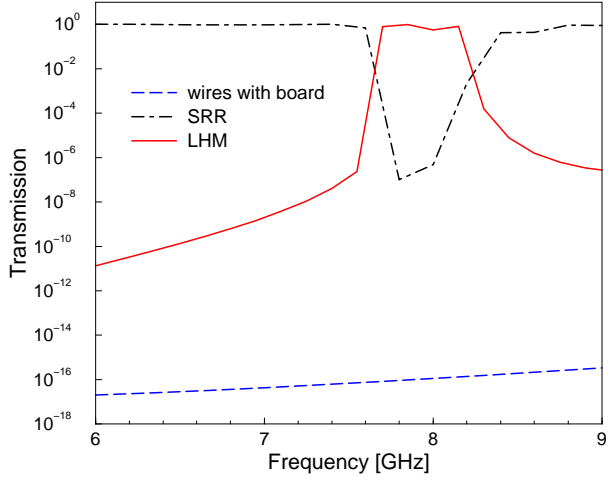,width=8cm}
\caption{ Transmission of electromagnetic waves  through arrays of SRRs,
wires, and
  through a  LHM with a real  permittivity for the metal $\ep= -3000$. The
dielectric
  constant of the board is $\eps_b=3.4$. The thickness of wire is $1\times
1$ mm.
Magnetic intensity is parallel to the SRR axis. For the other polarization,
no resonant gap was observed.
}
\end{figure}

\begin{figure}[t]
\epsfig{file=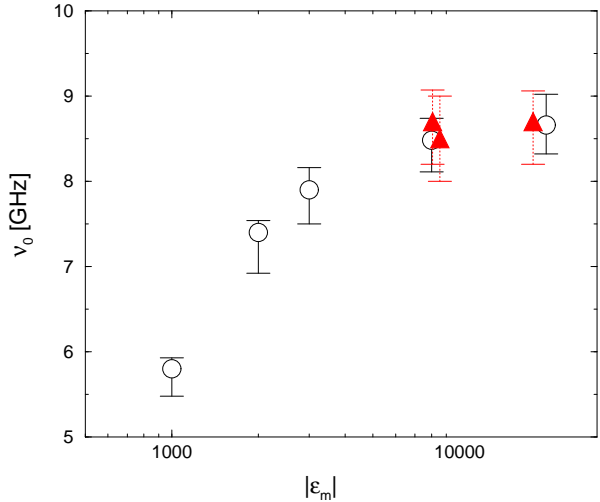,width=8cm}
\vspace*{2mm}
\caption{Dependence of the resonance frequency $\nu_0$
on the absolute value of the permittivity $|\ep|$ of metallic compounds.
Both real (open circles) and complex (solid triangles)
values of permittivity were considered. Error bars indicate the width of gap.
$\nu_0$ does not depend on $\eps$ for  $|\eps|\ge 10^4$.}
\end{figure}

\begin{figure}[t]
\epsfig{file=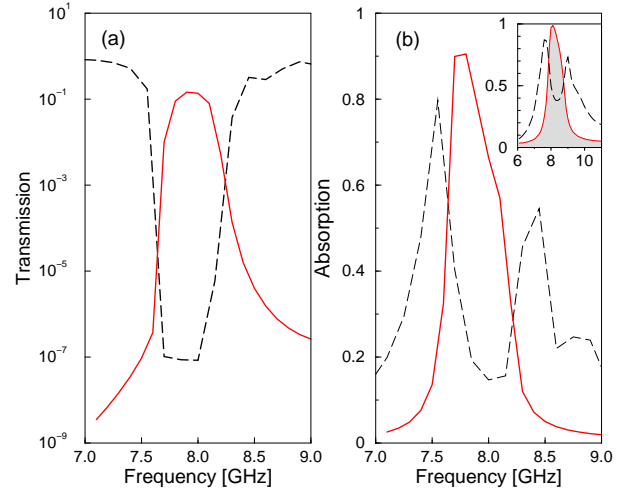,width=8cm}
\caption{ (a)  Transmission through arrays of SRRs (dashed line) and
through a LHM (solid
line) with a complex  permittivity for the metal. (b) Absorption versus
frequency for the
same system as shown in (a). Permittivity of metal is
$\eps=-3000+i~100$. Insert shows absorption in a composite system with a
frequency dependent
$\eps_{\rm eff}$ and $\mu_{\rm eff}$  given by Eqns. (\ref{eps_eff}) and
(\ref{mu_eff}).}
\end{figure}

\begin{figure}[t]
\epsfig{file=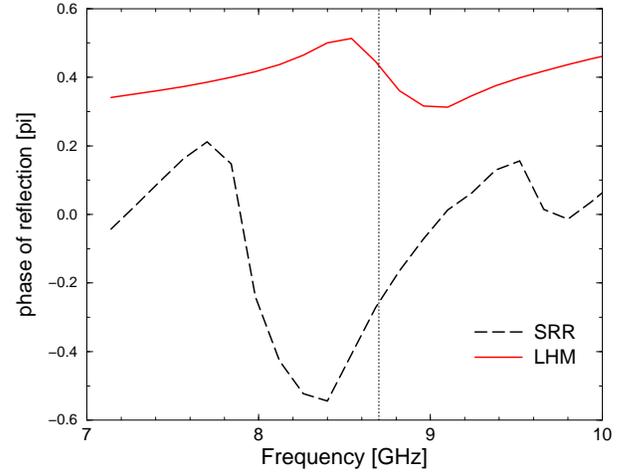,width=8cm}
\caption{Phase of reflection for both
SRR and LHM. $\ep=-1000 + i~18000$. The vertical dotted line indicates the
position of the center
  of the resonance gap $\nu_0$.}
\end{figure}

\begin{figure}
\epsfig{file=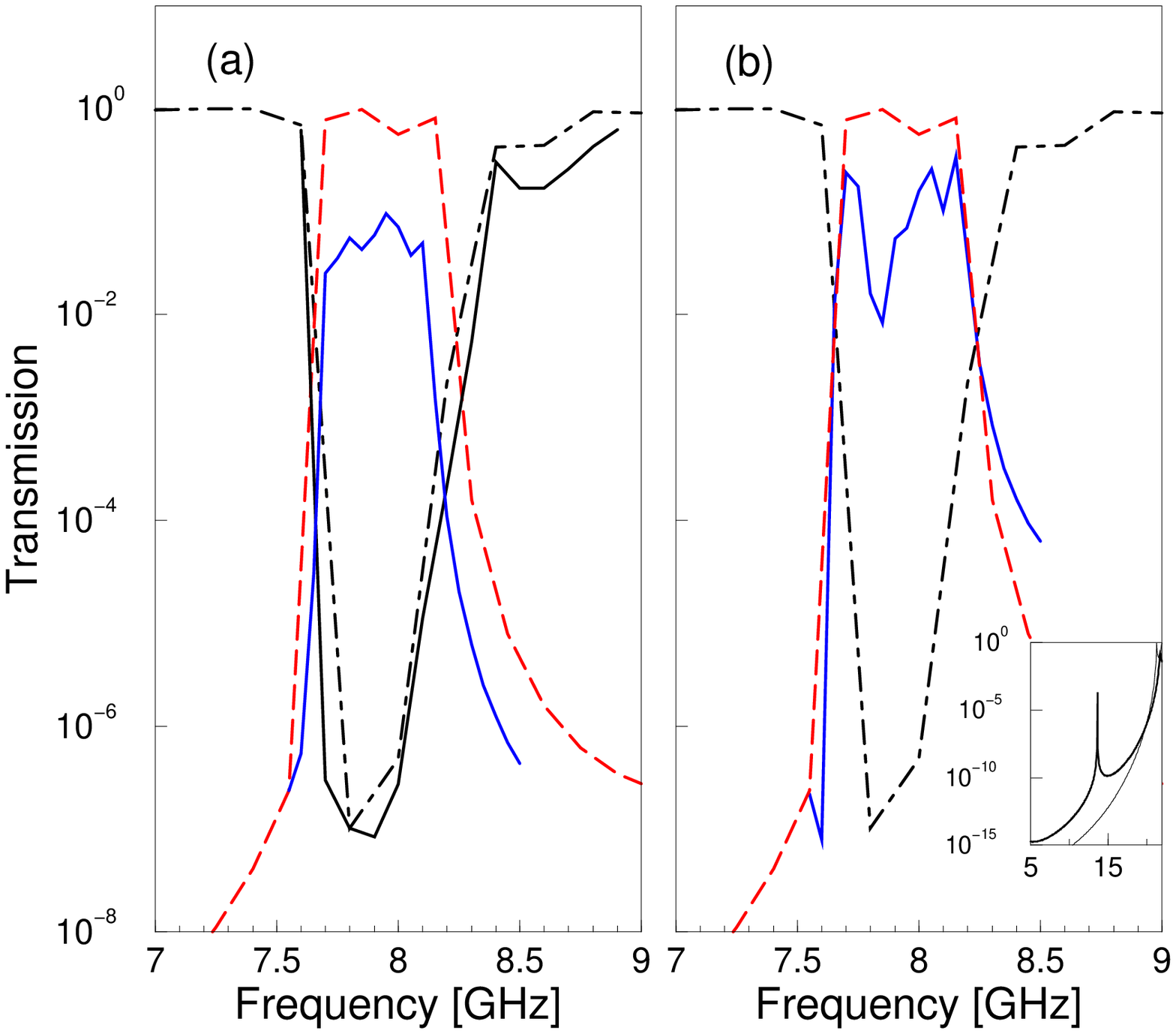,width=8cm}
\caption{Transmission of EM waves through SRRs (dot-dashed lines), 
and LHMs (dashed lines).
In (a) one SSR has been removed and its transmission is shown by a 
solid line with no appreciable difference from the
periodic case. In (b) one metallic wire has been removed and its 
transmission is shown as a solid line. In this case
there is a sharp dip in the transmission of the LHM with an impurity.
In all the cases $\epsilon_m=-3000$. For comparison, we also show as 
an inset in Fig. 6b the transmission of
waves through the lattice of wires only.  A sharp resonance peak with 
a frequency $\nu_{\rm resonance}\approx 13.65$
GHz appears when one wire is removed. Nine unit cells in the 
propagating direction are considered and the vacancies
were created in the middle  of the structure.}
\end{figure}.


\begin{thebibliography}{99}
\bibitem{cavity} In fact, due to the periodic boundary conditions
in the $x$ and $y$ directions, we have a planar cavity inside the system.
\end{thebibliography}
\end{document}